\documentclass[twoside]{article}
\usepackage{fleqn,espcrc2}
\usepackage{graphicx}
\usepackage{l3physics}

\newcommand{\GG}{\gamma^*\gamma^*}
\newcommand{\GGG}{\gamma\gamma}

\title{Double-tag events study with the L3 detector 
       at $\sqrt{s}=189 \GeV$}

\author{Pablo Achard
\address{University of Geneva, DPNC,
        24 quai E. Ansermet, \\ 
        1211 Gen\`eve 4, Switzerland}
	on behalf of the L3 Collaboration}
       
\begin{document}

\begin{abstract}
  A preliminary study of double tag events using the L3 detector at center of mass energy
  $\sqrt{s} \simeq ~189 \GeV$ has been performed. The cross-section of $\gamma ^* \gamma ^*$ 
  collisions is measured at average $ \langle Q^2 \rangle = 14.5~\rm{GeV}^2$. The results 
  are in agreement with predictions based on perturbative QCD, while the Quark Parton Model 
  alone is insufficient to describe the data. The measurements lie below the LO and above the
  NLO BFKL calculations.

\vspace{1pc}
\end{abstract}

\maketitle

\section{Introduction}

  \par
\begin{figure}[h]
  \begin{center}
  \includegraphics[width=15pc]{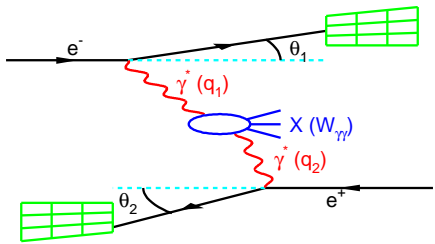}
     \vspace{-1pc}
  \caption[]{\it{ Kinematics of a double-tag event.
  \label{fig:kin}} }
\end{center}
\end{figure}

  In this paper we present an update of the analysis of double-tag  
  two-photon events:  $\epem \rightarrow \epem hadrons$ already published 
  for $\sqrt{s} \simeq 91 \GeV$ and $\sqrt{s} \simeq 183 \GeV$ ~\cite{paper_168}, 
  obtained at LEP with the L3 detector. The data, collected at centre-of-mass energy 
  $\sqrt{s} \simeq 189 \GeV$, correspond to an integrated luminosity of
  176~pb$^{-1}$. 
  Both scattered electrons\footnote{Electron stands for electron 
  or positron throughout this paper.} are detected in the small angle electromagnetic calorimeters
  (Fig.~\ref{fig:kin}). The virtuality of the two photons, $Q^2_1$ and $Q^2_2$, is 
  in the range of $3 \GeV^2 < Q^2_{1,2} < 37 \GeV^2$. 
  The centre-of-mass energy of the two virtual photons, $ \sqrt{\hat{s}}= W_{\GGG}$, ranges 
  from $3 \GeV$ to $75 \GeV$.

  \par
\begin{figure}[h]
  \begin{center}
  \includegraphics[width=15pc]{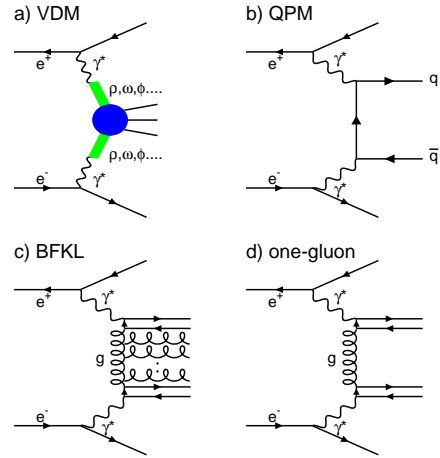}
     \vspace{-1pc}
  \caption[]{\it{ Diagrams for the a)~VDM, b)~QPM, c)~BFKL Pomeron and d)~one-gluon 
             exchange processes in $\GG$ interactions.
  \label{fig:feyn}} }
\end{center}
\end{figure}

  For $Q^2_1 \approx Q^2_2 \approx 0$ (untagged events)~\cite{l3tot}, the two-photon 
  cross-section, $\sigma_{\gamma\gamma}$, is dominated by vector-vector interactions, 
  VDM~(Fig.~\ref{fig:feyn}a).
  
  With increasing $Q^2$, the VDM process is suppressed by the vector meson form factor 
  and the Quark Parton Model process~(QPM), shown in Fig.~\ref{fig:feyn}b,
  becomes important. Single-tag two-photon events, where $Q^2_1 >> Q^2_2 \approx 0$, 
  are usually analysed within  
  the DIS formalism~\cite{L3F2} and a photon 
  structure function is introduced ({\it resolved photon}). 
  Since the highly virtual photon, unlike the proton, does not contain
  constituent quarks with an unknown density distribution, one may hope to have a complete 
  QCD calculation under particular
  kinematical conditions. In this formalism, used today by the
  Monte Carlo generators, one considers one
  or two resolved photons ({\it single or double resolved processes})
  to calculate the QCD leading order diagrams.
  
  \par
  An alternative QCD approach is based on the BFKL equation~\cite{bfkl}.
  Here the highly virtual two-photon process, with $Q^2_1\simeq Q^2_2$, is considered as the ``golden'' 
  process where the calculation can be verified without phenomenological inputs~\cite{gg2,gg1}.
  The $\gamma ^* \gamma^*$ interaction can  be seen as the  interaction of two $q \bar{q}$ pairs scattering 
  off each other via multiple gluon exchange~(Fig.~2c). 
  For $\ln {\hat{s}/Q ^2 } \approx 1$ 
  a diagram with one-gluon exchange could be sufficient and the 
  cross-section would be constant~(Fig.~2d). 
  In the limit of high energy, $\ln {\hat{s}/Q ^2 }>> 1$,
  the diagram of Fig.~2c is calculable  by the
  resummation of the large logarithms.
  In this scheme the cross-section for the collision of  two virtual
  photons is~\cite{gg2,gg1}:
  \begin{eqnarray}
    \sigma_{\GG} & = & \frac{\sigma_0}{\sqrt{Q_1^2 Q_2^2 Y}} \left(\frac{s}{s_0}\right)^{\alpha_P-1}.
  \end{eqnarray}
  Here 
  \begin{eqnarray}
    \sigma_0 & = & const ~~,~~
     s_0      =  \frac{\sqrt{Q_1^2 Q_2^2 }}{y_1y_2}   \nonumber \\
     y_i     & = & 1-(E_i/E_{b})\cos^2(\theta_i/2)   \nonumber  
  \end{eqnarray}
  where $E_{b}$ is the beam energy, $E_i$ and $\theta_i$ are the energy
  and polar angle of the scattered electrons and $\alpha_P$ is the ``hard Pomeron'' intercept. 
  The centre-of-mass energy of the two-photon system  
  is related to the $\epem$ centre-of-mass energy $s$ by $\hat{s} = W^2_{\GGG} \approx s y_1y_2$.
  In leading order one has $\alpha_P-1  = (4\ln2) N_c\alpha_s /\pi$  
  where  $N_c$ is the number of colours. Using $N_c = 3$ and
  $\alpha_s =0.2$, one obtains $\alpha_P -1 \simeq 0.53$~\cite{gg2,gg1}; in the 
  next-to-leading order the BFKL contribution is calculated to be smaller, 
  $\alpha_P -1 \leq 0.17$~\cite{gg3}.

  \par
  Double-tag interactions have been measured in previous experiments~\cite{oldggs} 
  at lower values of $Q^2$ and $W_{\gamma\gamma}$. For comparison with the prediction of the 
  BFKL models, the cross-sections will be given as a function of the variable $Y=\ln{(s/s_0)}$ 
  instead 
  of $W_{\gamma\gamma}$ as used in Ref.~\cite{l3tot,oldggs}. The advantage 
  in using this variable is that $Y$ is independent of the beam energy :
  \begin{eqnarray}
     Y  \approx ln\left(\frac{(q_1+q_2)^2}{\sqrt{q_1^2q_2^2}}\right)
	  \approx  ln\left(\frac{q_1^2+2q_1.q_2+q_2^2}{q_1q_2}\right) \nonumber 
  \end{eqnarray}
   with $q_{1,2}^2 = - 2 \, \sqrt{s} \, E_{tag_{1,2}} \, (1-cos~\theta_{1,2})$.\\
   As ~$E_{tag_{1,2}} \simeq \sqrt{s}/2$ ~and~ $m_e \simeq 0$: \\
  \begin{eqnarray}
     Y\approx ln\left(\frac{2-cos ~\theta_{\gamma_1\gamma_2}}{2\sqrt{(1-cos~\theta_1)(1-cos~\theta_2)}}\right)
  \end{eqnarray}
   where $\theta_{\gamma_1\gamma_2}$ is the angle between the two photons and $\theta_{1,2}$ are 
   the tagging angles.
   
\section{Monte Carlo Generators}  
  \par
  The Monte Carlo generators used in this analysis are JAMVG~\cite{verm} which generates 
  events with the matrix element of Figure~\ref{fig:feyn}b and  PHOJET~\cite{pho}
  which gives a good description of the single-tag events~\cite{L3F2} and uses the 
  GRV-LO~\cite{grv} parton density in the photon to initiate QCD processes.
  
  \par
  The dominant backgrounds are $\epem \rightarrow \epem \tau^+\tau^-$,
  simulated by JAMVG~\cite{verm}, and single-tag two-photon hadronic events, where a hadron is misidentified as a scattered
  electron.
  The contamination by annihilation processes is simulated by PYTHIA~\cite{pythia}
  ($\epem \rightarrow hadrons$),
  KORALZ~\cite{kora}~($\epem \rightarrow \tau^+\tau^-$) and KORALW~\cite{korw}~($\epem \rightarrow \mathrm{W^+W^-}$).

  \par
  All Monte Carlo events are passed through a full detector simulation using the GEANT~\cite{GEANT} and 
  the GEISHA~\cite{GEISHA} programs and are reconstructed in the same way as the data.
  
\section{Data Analysis}
\subsection{Event Selection}

  \par
    A detailed description of the
    L3 detector is given in Ref.~\cite{l3_000,l3det:bro}.
    The two-photon hadronic events are mainly triggered by two independent triggers:
    the central track~\cite{LVT} and the single and double tag energy~\cite{LVE} triggers. 
The total trigger inefficiency of the selected events is 
    less than 1\%.

\begin{figure}[ht]

   \begin{center}
     \includegraphics[width=15pc]{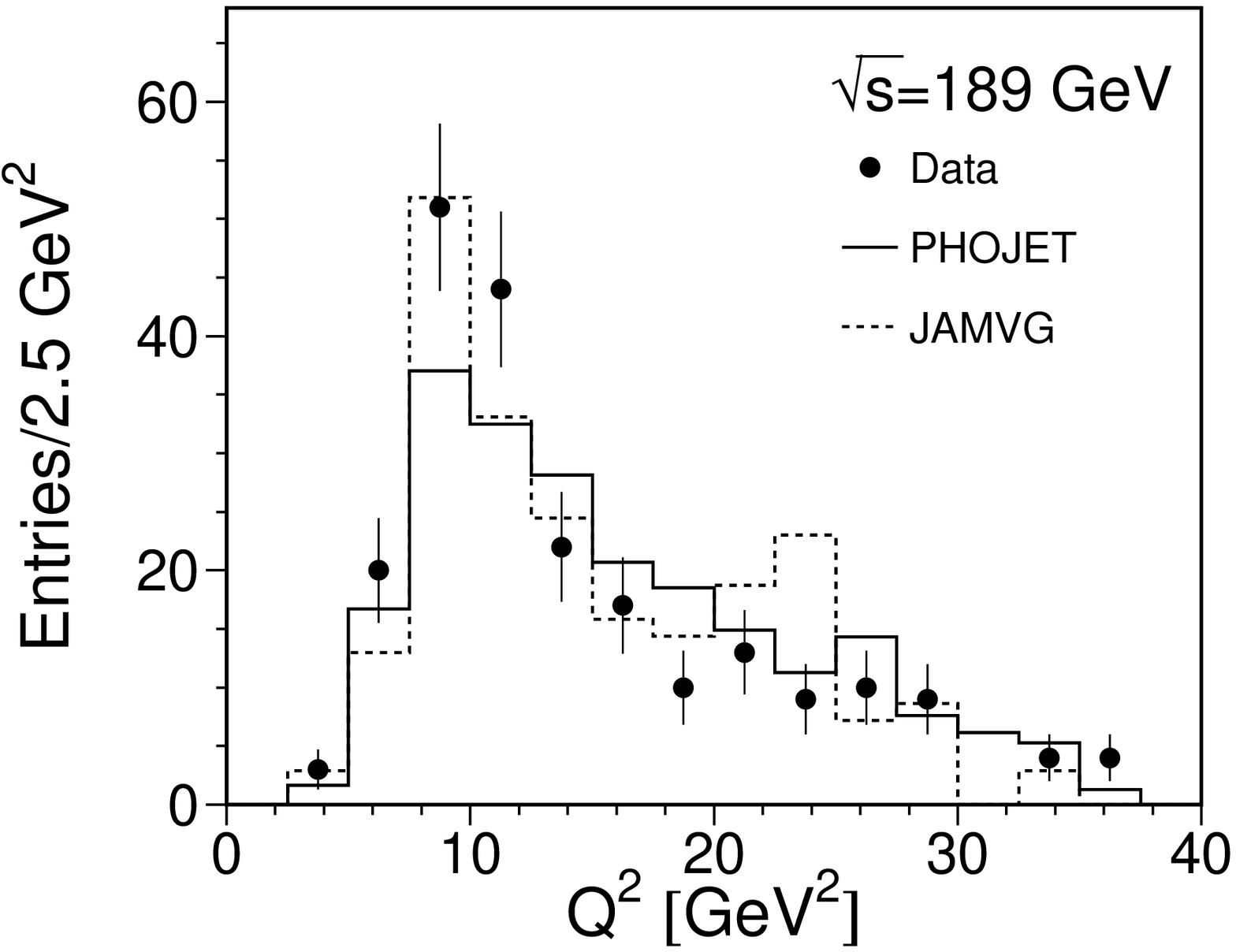}
     \includegraphics[width=15pc]{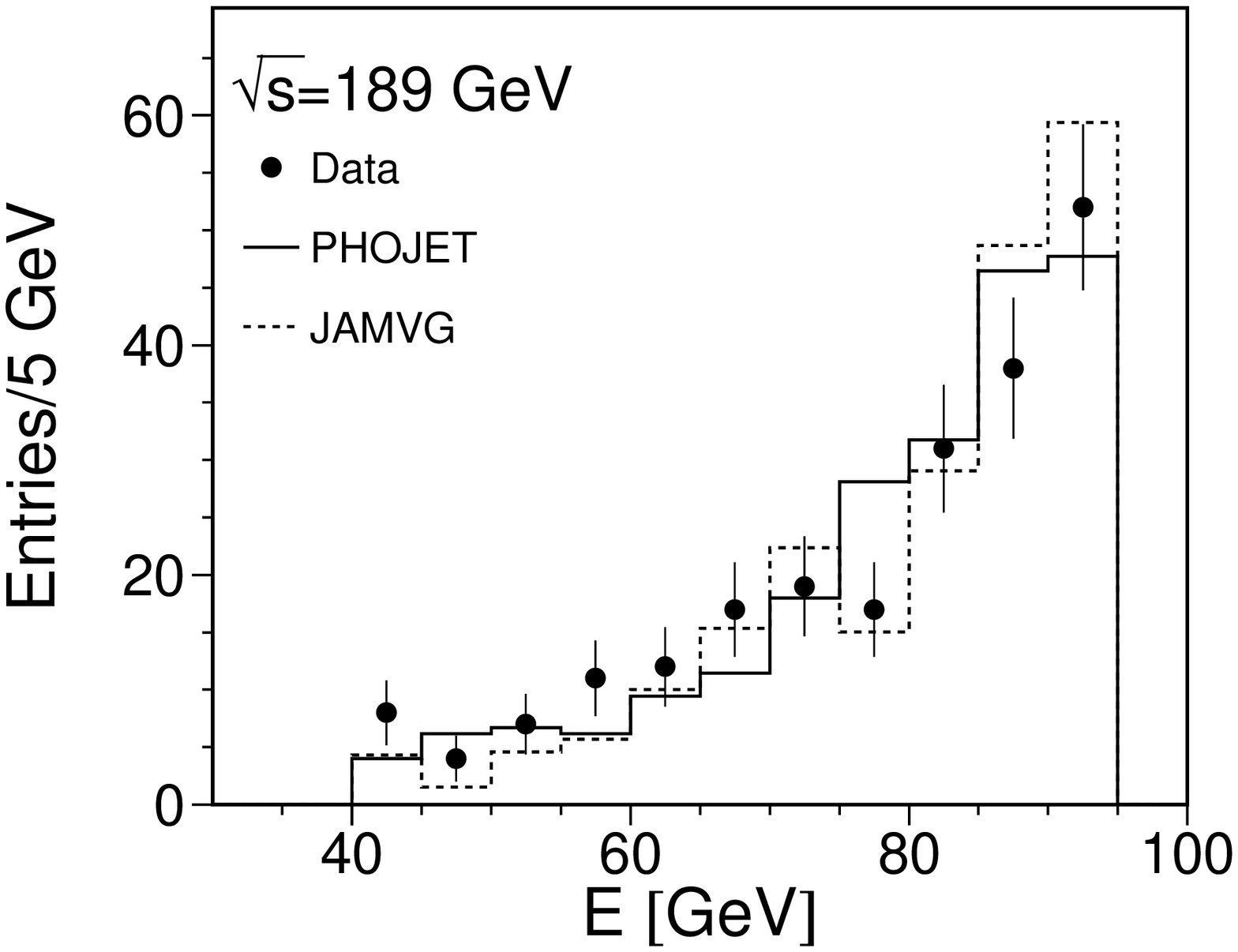}
     \vspace{-3pc}
   \end{center}
  \caption[]{\it{ Distributions of $Q^2$~and energy of the scattered electrons.
           The data is compared to the 
           Monte Carlo predictions, normalised to the number of data events 	   
           There are two entries per event.
  \label{fig:q2} }}
\end{figure}

    Two-photon hadronic event candidates, $\epem \rightarrow \epem hadrons$, are selected using 
    the following cuts:
    \begin{itemize}
     \item There must be two identified electrons, forward and backward~(double-tag), in the small angle electromagnetic
           calorimeters. Each electron is identified as the highest energy cluster in one of the 
           calorimeters, with energy greater than $40 \GeV$. The polar angle of the two tagged electrons 
	   has to be in the range $30~\mathrm{mrad}<\theta_1 < 66$~mrad and 
	   $30~\mathrm{mrad}<\pi-\theta_2 < 66$~mrad.
     \item The number of tracks, measured in the polar angle region $20^\circ<\theta < 160^\circ$, must be 
           greater than two. The tracks are required to have a 
           transverse momentum, $p_t$, greater than $100 \MeV$ and a distance of closest approach in the transverse plane to the
           interaction vertex smaller than 10~mm.
     \item The value of $Y=\ln{(s/s_0)}$ is required to be in the range $2 \leq Y \leq 6$.
    \end{itemize}
    After these cuts, 108 events are selected with an estimated background of 14 events  
    from  $\epem \rightarrow \epem \tau^+\tau^-$ and misidentified single-tag events. 
    The contamination from annihilation processes is negligible.

\begin{figure}[t]
   \begin{center}
     \includegraphics[width=15pc]{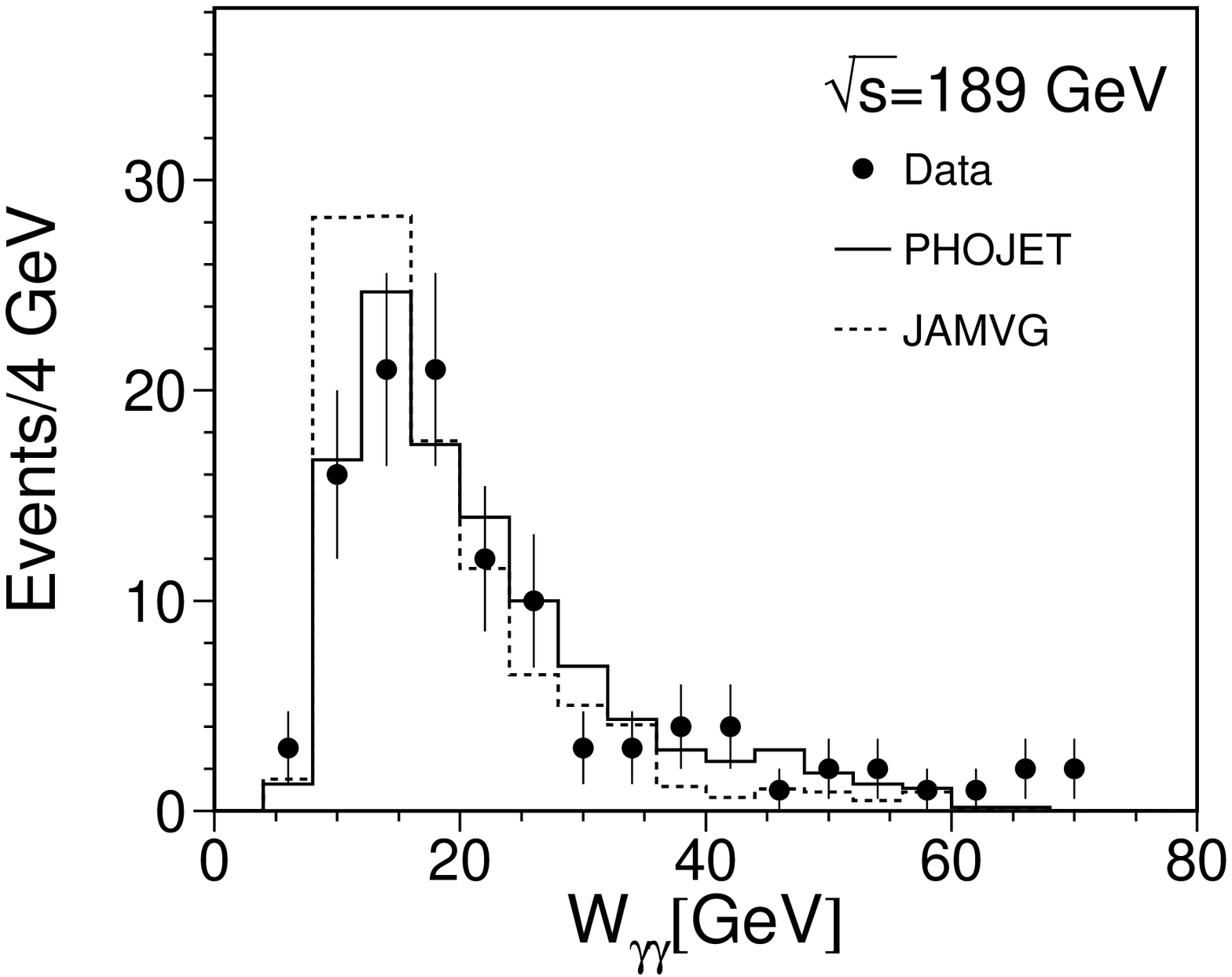}
    \includegraphics[width=15pc]{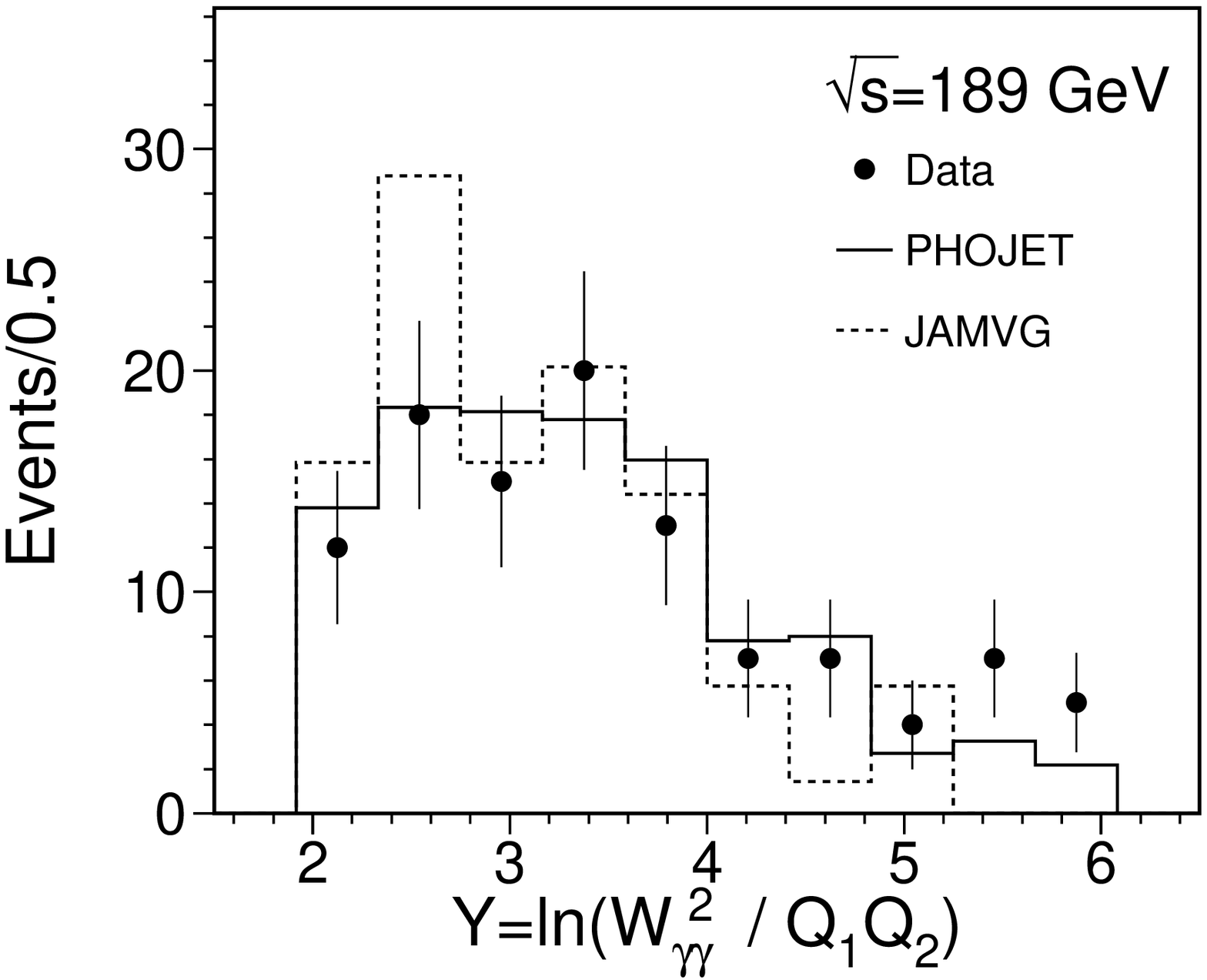}
     \vspace{-3pc}
   \end{center}
  \caption[]{\it{ Distributions of the two-photon mass, $W_{\gamma\gamma}$,~and the variable
             $Y$. The data are compared to the Monte Carlo predictions, normalised to the 
	     number of data events.
  \label{fig:wtru}} }
\end{figure}

    \begin{table*}[hbt]          
      \begin{center} 
              \caption[]{\it{ The differential cross-section, 
	            d$\sigma (\epem \rightarrow \epem hadrons)$/d$Y$ in picobarn measured
                   in the kinematic region defined in the text, at $\sqrt{s} \simeq 189 \GeV$.
		   The predictions of the PHOJET and the JAMVG Monte Carlo are also listed.
		   The first error is statistical and the second is systematic.
       \label{tab:cro} }}
        \begin{tabular}{ccccc}
          \hline           
               & DATA & PHOJET    &   DATA  &  JAMVG \\
         &~~\small corrected with PHOJET~~&          & ~~\small corrected with  JAMVG ~~& \\ \hline
         $\Delta Y$ & $\mathrm{d\sigma /d}Y$ & $\mathrm{d\sigma /d}Y$ & $\mathrm{d\sigma /d}Y$ & $\mathrm{d\sigma /d}Y$  \\ \hline
      $2-3$    &  $0.39 \pm 0.06 \pm 0.06$  & 0.32  &  $0.39 \pm 0.07 \pm 0.06 $ & 0.27 \\ 
      $3-4$    &  $0.24 \pm 0.04 \pm 0.04$  & 0.21  &  $0.22 \pm 0.04 \pm 0.03 $ & 0.11 \\ 
      $4-5$    &  $0.11 \pm 0.03 \pm 0.02$  & 0.09  &  $0.08 \pm 0.02 \pm 0.01 $ & 0.03 \\ 
      $5-6$    &  $0.08 \pm 0.03 \pm 0.01$  & 0.04  &  $0.07 \pm 0.02 \pm 0.01 $ & 0.01 \\ \hline

        \end{tabular}
      \end{center}
    \end{table*}
  \par
    Contrary to the case of untagged or single-tag events the $W_{\gamma\gamma}$ measurement 
    does not rely on the $W_{vis}$ measurement. The analysis is, therefore, less dependent 
    on the Monte Carlo modelling of the structure of the final state.
    The distributions 
    of $Q^2_i$ and $E_i$ of the two scattered electrons are shown in Fig.~\ref{fig:q2}. 
    In Fig.~\ref{fig:wtru}, the distributions of $W_{\gamma\gamma}$ and $Y$
    are presented. The variable $W_{\gamma\gamma}$ is calculated using the kinematics of the two scattered electrons,
    taking advantage of the good resolution of the energy of scattered electrons~(about 1.3\%~\cite{l3det:bro}). 
    The $W_{\gamma\gamma}$ resolution is about 18\% at $W_{\gamma\gamma}/\sqrt{s} = 0.08$ and about 6\% 
    at $W_{\gamma\gamma}/\sqrt{s} = 0.2$.  PHOJET gives a reasonable description of the shape of data. The 
    absolute normalisation of PHOJET is about 20\% too low.

%%%%%%%%%%%%%%%%%%%%%%%%%%%%%%%%%%%%%
% DOUBLE TAG CROSS SECTION          %  
%%%%%%%%%%%%%%%%%%%%%%%%%%%%%%%%%%%%%
  
\subsection{Double-tag cross-section}

  \par
    The cross-sections are measured in the kinematic region limited by:
    \begin{itemize}
     \item $E_{1,2} > 30 \GeV$, $30~\mathrm{mrad} < \theta_{1} < 66$~mrad and 
     $30~\mathrm{mrad} < \pi-\theta_{2} < 66$~mrad.
     \item $2 \leq Y \leq 6$.
    \end{itemize}
   The data are corrected for efficiency and acceptance with PHOJET.
    The correction factors vary from about 30\% at low values of $Y$ to about 80\%
    at high values of $Y$. The correction factors calculated with JAMVG
    are similar. The differential cross-sections  
    $\mathrm{d}\sigma (\epem \rightarrow \epem hadrons) /\mathrm{d}Y$ are
    measured in four $\Delta Y$ intervals and are listed in Table~\ref{tab:cro}.

  \par
    The systematic error due to the selection cuts is 5\%, estimated by 
    varying the cuts. The uncertainty on the cross-section from the background estimation of 
    single-tag events is 11\%. The uncertainty due to the acceptance correction is  9\%. 
    The different systematic uncertainties are added in quadrature to give
    the total systematic error listed in Table~\ref{tab:cro}.

  \par
   ~~\\
   \vspace{-1cm}
	
\begin{figure}[h]
  \begin{center}
  \includegraphics[width=15pc]{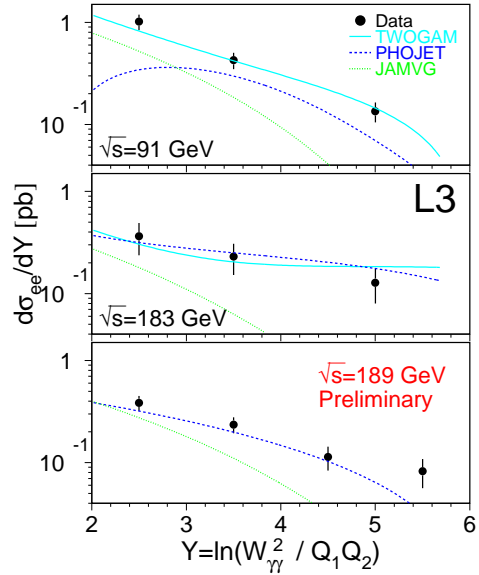}
     \vspace{-2pc}
  \caption[]{\it{ The cross-section of $\epem \rightarrow \epem hadrons$ as a function of $Y$ in the
             kinematical region defined in the text at $\sqrt{s} \simeq 91, 
	     183~\mathrm{and}~189\GeV$ compared to the predictions of the Monte Carlo 
	     models~\cite{pho,two,verm}.
  \label{fig:cro} }}
\end{center}
\end{figure}

  \par
    As can be seen in Table~\ref{tab:cro} and in Fig.~\ref{fig:cro}, the data are in reasonable
    agreement with the predictions of the  QCD Monte Carlo model implemented in PHOJET
    \footnote{The poor agreement at low Y for $\sqrt{s}=91 \GeV$ is well understood (suppression 
    of direct contribution due to a too sever $p_T$ cut)}, 
    whereas the QPM cross-section estimated with JAMVG, is not sufficient to describe the data.   
	
  \par
    From the measurement of the $\epem \rightarrow \epem hadrons$ cross-section, 
    $\sigma_{\mathrm{ee}}$, we extract the two-photon cross-section, $\sigma_{\gamma^*\gamma^*}$, 
    by using only the transverse photon luminosity function~\cite{budnev,lumi},
    $\sigma_{\mathrm{ee}} = L_{TT} \cdot  \sigma_{\GG}$.
    This measurement gives an effective cross-section containing contributions from 
    transverse~($T$) and longitudinal~($L$) photon polarisations:
    \begin{eqnarray}
      \sigma_{\GG} =  \sigma_{TT} + \epsilon_1\sigma_{TL} + \epsilon_2\sigma_{LT} + \epsilon_1 \epsilon_2 \sigma_{LL}
    \end{eqnarray}
    \vspace{-2pc}
    \begin{eqnarray}
      \epsilon_i = \frac { L_{L}} {L_{T}} = \frac { 2 (1-y_i)} {1+(1-y_i)^2}  \nonumber
    \end{eqnarray}
    where $\epsilon_i$ is the ratio of transverse and longitudinal photon luminosity functions. 
    In the present kinematical region the values of $\epsilon_i$
    are  greater than 0.97, but the values of $\sigma_{TL}$, $\sigma_{LT}$ and
    $\sigma_{LL}$ are expected to be small~\cite{gg2}.
%    The values of the two-photon cross-sections are given in Table~\ref{tab:ggcro}.

  \begin{table}[h]
    \begin{center} 
     \caption[]{\it{ The two-photon cross-section, $\sigma_{\GG}$ in nanobarn without and
     		after subtraction of the QPM contribution, as a function of $Y$ 
     		at $\sqrt{s} \simeq 189 \GeV$. The first error is statistical 
		and the second is systematic.
     \label{tab:ggcro}} }
     \begin{tabular}{ccc}
       \hline
                  &                                  &   \\  
      $\Delta Y$  &$\sigma_{\gamma^*\gamma^*}^{TOT}$   &
      $\sigma_{\gamma^*\gamma^*}^{TOT - QPM}$  \\  
                  &                 &                \\ \hline
      $~~2-3$       &~~~$7.5 \pm  1.3 \pm 1.1$   ~~~&$2.2 \pm 0.4 \pm 0.3$ ~~  \\ 
      $~~3-4$       &~~~$7.3 \pm  1.3 \pm 1.1$   ~~~&$3.8 \pm 0.6 \pm 0.6$ ~~  \\ 
      $~~4-5$       &~~~$5.5 \pm  1.5 \pm 0.8$   ~~~&$4.2 \pm 1.1 \pm 0.6$ ~~  \\ 
      $~~5-6$       &~~~$7.4 \pm  2.3 \pm 1.1$   ~~~&$6.7 \pm 2.1 \pm 1.0$ ~~  \\ 
     \hline	 
	  
     \end{tabular}
   \end{center}
  \end{table}

  \par
\begin{figure}[h]
  \begin{center}
  \includegraphics[width=15pc]{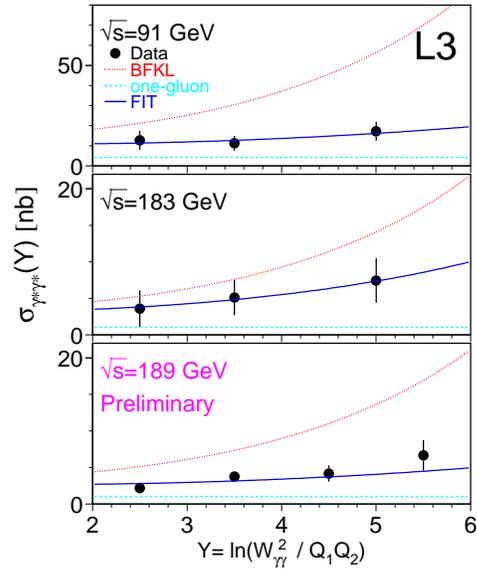}
     \vspace{-2pc}
  \caption[]{\it{ Two-photon cross-sections, $\sigma_{\gamma^*\gamma^*}$, after subtraction of the QPM contribution
             at $\sqrt{s} \simeq 91 \GeV$ ($\langle Q^2 \rangle = 3.5 \GeV^2$), 
             $\sqrt{s} \simeq 183 \GeV$~($\langle Q^2 \rangle = 14 \GeV^2$)and
	     $\sqrt{s} \simeq 189 \GeV$~($\langle Q^2 \rangle = 14.5 \GeV^2$). 
	     The data are compared to the predictions of the LO BFKL calculation
             and of the one-gluon exchange diagram. The continuous line is a fit to the data with Eq.~1 by leaving $\alpha_P$
             as a free parameter.
  \label{fig:ggcro}} }
\end{center}
\end{figure}

    In Fig.~\ref{fig:ggcro} we show $\sigma_{\GG}$, after subtraction of the QPM contribution given 
    in Table~\ref{tab:ggcro}, as a function of $Y$.
    Using an average value of $Q^2$, $\langle Q^2 \rangle = 14.5 \GeV^2$, 
    we calculate the one-gluon exchange contribution with the asymptotic 
    formula~(Eq.~10.2 of Ref.~\cite{gg2}). The expectations are below
    the data. The leading order expectations of the BFKL model~(Eq.~4.19 of Ref.~\cite{gg2}), 
    shown as a dotted line in Fig.~\ref{fig:ggcro}, are too high. By leaving $\alpha_P$ as 
    a free parameter, a fit to the data, taking into account the statistical errors, yields:  
    \begin{eqnarray}
       \alpha_P - 1 &= 0.29 \pm 0.03~~.
    \end{eqnarray}
   \par
    The results are shown in Fig.~\ref{fig:ggcro} as a continuous line. In the same figure
    are also shown our previous results~\cite{paper_168} obtained at 
    $\sqrt{s} \simeq 91  \GeV$  and  $\sqrt{s} \simeq 183 \GeV$ . A combined fit of the three 
    sets of data leaving $\alpha_P$ as a free parameter gives:
    \begin{eqnarray}
       \alpha_P - 1 = 0.29 \pm 0.025~~
    \end{eqnarray}       
       with ~~ $\chi^2 / d.o.f. = 7 / 10 ~~(C.L.=0.73)$.

    \par
     But if one fixes $\alpha_P$ to its LO and NLO calculated values, one finds respectively:
    \begin{eqnarray}
     \chi ^2_{\, LO-BFKL} / d.o.f. = 263 / 10 ~~(C.L.<10^{-16})\nonumber
    \end{eqnarray}
    \begin{eqnarray}
    \vspace{-4pc}
     \chi ^2_{\, NLO-BFKL} / d.o.f. = 21 / 10  ~~(C.L.=0.021).\nonumber
    \end{eqnarray}

\section{Conclusions}
  \par
    The cross-sections of double-tag $\epem \rightarrow \epem hadrons$
    events is measured at $\sqrt{s} \simeq 189 \GeV$. The events are well descibed by 
    the PHOJET Monte Carlo model which uses the GRV-LO parton density in the photon  
    and leading order perturbative QCD. A combined fit of our results obtained at 
    $\sqrt{s} \simeq 91, 183~\mathrm{and}~189 \GeV$ gives a value of $\alpha_P-1$ 
    smaller than expected from the LO BFKL calculations and higher than the value predicted by
    NLO BFKL calculations.
         
\section*{Acknowledgements}
   \par
    I would really like to thank Maneesh Wadhwa for providing me the results
    and some help, 
    and all the L3 $\GGG$ group for discussions and support.\\
    Thanks also to the organizers of this interesting conference.

\end{document}